\documentclass[a4,12pt]{article}
\usepackage{graphics}
\usepackage[usenames]{color}
\usepackage[dvips]{epsfig}
\usepackage{mathrsfs}
\usepackage{amsmath,amssymb}
\newcommand{\be}{\begin{equation}}
\newcommand{\ee}{\end{equation}}
\newcommand{\bea}{\begin{eqnarray}}
\newcommand{\eea}{\end{eqnarray}}
\newcommand{\bean}{\begin{eqnarray*}}
\newcommand{\eean}{\end{eqnarray*}}
\newcommand{\ba}{\begin{array}}
\newcommand{\ea}{\end{array}}
\newcommand{\non}{\nonumber}
\newcommand{\bc}{\begin{center}}
\newcommand{\ec}{\end{center}}
\newcommand{\bi}{\begin{itemize}}
\newcommand{\ei}{\end{itemize}}

\newcommand{\norsl}{\normalsize\sl}
\newcommand{\norsc}{\normalsize\sc}

\newcommand{\ra}{\rightarrow}

\newcommand{\aqt}{{\mathscr{A}_{TT}(Q_T)}}
\newcommand{\aqtn}{\mathscr{A}_{TT}}

\begin{document}

\title{Double-spin asymmetries for small-$Q_T$ 
Drell-Yan pair production
in transversely polarized $p \bar{p}$ collisions
%
}   

\author{
{\norsc  Hiroyuki Kawamura${}^a$, Jiro Kodaira${}^b\footnote{Deceased.}
$ and Kazuhiro Tanaka${}^c$}\\
\\
\norsl  ${}^a$ Radiation Laboratory, RIKEN, Wako 351-0198, 
Japan \\
\norsl  ${}^b$ Theory Division, KEK, Tsukuba 305-0801, Japan
\\
\norsl  ${}^c$ Department of Physics, Juntendo University,
    Inba, Chiba 270-1695, Japan}
\date{}
\maketitle

\begin{abstract}
   We discuss the Drell-Yan process at a measured transverse-momentum $Q_T$
   of the produced lepton pair in collisions of transversely polarized
   protons and antiprotons, to be observed at the proposed spin experiments at GSI.
   The large logarithmic contributions from multiple soft gluon emission, 
   accompanying the Drell-Yan mechanism at small $Q_T$, are resummed to all orders 
   in QCD perturbation theory up to next-to-leading logarithmic (NLL) accuracy.
Numerical evaluation shows the
impact of the NLL as well as LL 
effect on the dilepton $Q_T$ spectra.
For the corresponding $Q_T$-dependent spin asymmetry $\aqt$, 
the LL effect gives significant 
modification
while
the NLL effect is marginal,
%
leading to QCD prediction that
$\aqt$ at GSI is flat at small and moderate $Q_T$
and almost 
equals the conventional asymmetry $A_{TT}$ 
associated with the $Q_T$-integrated cross sections.
   This flat behavior in turn allows us to use analytic saddle-point 
   evaluation of the 
resummation formula in the limit $Q_T \rightarrow 0$,
   not only to obtain quantitative estimate of $\aqt$, but also to clarify mechanisms 
   behind 
the relation 
$\aqt \simeq A_{TT}$ characteristic of $p\bar{p}$ collisions at GSI.
\end{abstract}

\newpage
Recently there has been much progress to unveil the last unknown parton 
distribution of nucleon at the leading twist,  
the transversity distribution $\delta q(x)$ \cite{Ralston:1979ys,BDR:02}:
for example, the calculation of the lowest two moments of $\delta q(x)$ 
by lattice QCD simulation is updated \cite{SFQCD:05}, and 
the first global fit of $\delta q(x)$ is given \cite{Anselmino:07}
using the semi-inclusive deep inelastic scattering (SIDIS) data, in combination 
with the $e^{+}e^{-}$ data for the associated (Collins) fragmentation function.
These results indicate that the transversity distributions for $u$ and 
$d$ quarks are sizable, but are fairly small compared with their positivity bound
(Soffer bound) \cite{Soffer:95}; still, these results are subject to 
certain assumptions and uncertainties, and it is important 
to proceed further toward the determination of the transversity.

%
As is well-known~\cite{Ralston:1979ys,BDR:02},
the transversely polarized Drell-Yan (tDY) process provides another 
promising way to access $\delta q(x)$.
The measurement of the tDY cross 
section is proposed in the future experiments at GSI \cite{PAX:05},
where the $p\bar{p}$ collisions at moderate energy mainly probes 
the products of two quark transversity-distributions, 
$\delta q(x_1) \delta q(x_2)$, in the ``valence region''.
%
%
%
%
%
%
The corresponding double-spin asymmetries estimated at the leading order (LO) in QCD 
are large enough to be measured at GSI \cite{Anselmino:2004ki}, 
in contrast to 
the complementary case of tDY in $pp$ collisions at RHIC,
where the asymmetries are predicted to be rather small because 
$\delta q(x_1) \delta \bar{q}(x_2)+(1\leftrightarrow 2)$ 
is probed in the ``sea-quark region'' 
(see \cite{MSSV:98,KKT:07,KKT:07-2}).
%
%
As for the higher-order QCD corrections for tDY in $p\bar{p}$ collisions, 
the NLO corrections have been studied recently~\cite{BCCGR:06};
also, the ``threshold resummation'' has been applied in order to sum up, 
to all orders in $\alpha_s$,
the soft-gluon emission contributions that are logarithmically enhanced 
near the threshold of the partonic process \cite{SSVY:05}.
The effects of these QCD corrections are 
small for the double transverse-spin asymmetries in $p\bar{p}$ collisions 
at the kinematical regions corresponding to the GSI experiments \cite{BCCGR:06,SSVY:05},
suggesting that the large LO asymmetries obtained in \cite{Anselmino:2004ki} 
are rather robust.

All the above previous studies of tDY in $p\bar{p}$ collisions
considered the case in which 
the transverse-momentum $Q_T$ of the produced lepton pair are unobserved.
Experimentally, however, 
the bulk of events is produced in the small $Q_T$ region.
%
%
Therefore, it is desirable to develop theoretical predictions of tDY 
at a measured $Q_T$ in $p\bar{p}$ collisions
for the detailed comparison with the data in the future GSI experiments.
When $Q_T\ll Q$ with $Q$ the dilepton mass, the fixed-order
perturbation theory breaks down due to the appearance of large logarithms 
$\ln (Q^2/Q_T^2 )$ multiplying $\alpha_s$, 
and we have to deal with the relevant DY cross sections at small $Q_T$ 
in an all-order resummation framework in QCD.
The corresponding ``$Q_T$-resummation'' has formally some resemblance 
to the threshold resummation mentioned above, but represents other 
contributions associated with different ``edge region''
of phase space.
The $Q_T$-resummation for tDY has been formulated recently 
by the present authors \cite{KKST:06,KKT:07,KKT:07-2}, summing the corresponding 
large logarithms up to next-to-leading logarithmic (NLL) accuracy,
in the context of the $pp$ collisions at RHIC and J-PARC.
There we demonstrated that profound modifications arise in the DY production 
at small $Q_T$,
driven by partonic mechanism that induces the large logarithmic contributions;
in particular, the interplay between the Sudakov factor resumming
multiple soft gluon emission and the DGLAP evolutions of parton distributions 
yields the ``amplification'' of the double transverse-spin asymmetries 
in the small $Q_T$ region at RHIC as well as J-PARC kinematics, 
resulting in their values larger than the conventional (fixed-order) NLO
asymmetries \cite{MSSV:98} calculated in the $Q_T$-unobserved case.
In this Letter we apply our $Q_T$-resummation formalism to the case of 
$p\bar{p}$ collisions, and study 
the corresponding double transverse-spin asymmetries in the small $Q_T$ region 
for tDY foreseen at GSI.

Detailed derivation of the tDY cross sections in our $Q_T$ resummation 
formalism is given in \cite{KKT:07-2}
for collisions of spin-1/2 hadrons $h_1$ and $h_2$, i.e., 
$h_1 h_2 \rightarrow  l^+l^-X$, and 
the corresponding results for $p\bar{p}$ collisions
can be obtained by trivial substitutions. 
The spin-dependent 
($\Delta_T d \sigma \equiv (d\sigma^{\uparrow\uparrow}-d\sigma^{\uparrow\downarrow})/2$)
and spin-independent 
($d \sigma \equiv (d\sigma^{\uparrow\uparrow}+d\sigma^{\uparrow\downarrow})/2$)
parts of the 
differential cross section 
are expressed as 
\be
\frac{\left(\Delta_T  \right)d\sigma}{dQ^2dQ_T^2dyd\phi} =
\left(\cos( 2\phi)/2\right)
\frac{2\alpha^2}{3\, N_c\, S\, Q^2}
\biggl[ \left(\Delta_T\right) \tilde{X}^{\rm NLL} (Q_T^2 , Q^2,y)
+ \left(\Delta_T\right) \tilde{Y} (Q_T^2 , Q^2,y)\biggr] \ .
\label{NLL+LO}
\ee
Here and below we follow the convention of \cite{KKT:07,KKT:07-2} for the basic 
quantities entering the formulae:
$\sqrt{S}$ and $y$ are the total energy and dilepton's rapidity 
in the proton-antiproton CM system, and 
the prefactor $\cos(2\phi)/2$ specific to the spin-dependent part 
shows the characteristic dependence~\cite{Ralston:1979ys} on 
the azimuthal angle $\phi$ 
of one of the outgoing leptons with respect to the incoming nucleon's spin axis.
The first term ($\Delta_T \tilde{X}^{\rm NLL}$
or $\tilde{X}^{\rm NLL}$)
gives the dominant contribution when $Q_T \ll Q$, containing 
all the logarithmically-enhanced contributions
$\alpha_s^n\ln^m(Q^2/Q_T^2)/Q_T^2$
and, at the NLL accuracy, has to be evaluated by resumming the first three 
towers ($m=2n-1, 2n-2, 2n-3$) of these large logarithmic
contributions to all orders 
in $\alpha_s$.
The second term ($\Delta_T\tilde{Y}$ 
or $\tilde{Y}$)
is free of such contributions,
and can be computed by fixed-order truncation of the perturbation theory. 

The NLL resummed component $(\Delta_T )\tilde{X}^{\rm NLL}$
is 
obtained through various kinds of 
elaboration~\cite{KKST:06,KKT:07,KKT:07-2,BCDeG:03,LKSV:01}
of the Collins-Soper-Sterman (CSS) resummation formalism~\cite{CSS:85}.
Introducing the impact parameter $b$ space, which is conjugate to the $Q_T$ space,
we have\footnote{
In this paper, we set $\mu_R=\mu_F =Q$ for the renormalization and factorization scales
$\mu_R$ and $\mu_F$.
} 
%
\bea
\Delta_T \tilde{X}^{\rm NLL} (Q_T^2, Q^2, y)
&&\!\!\!\!\!\!\!=
\int_{\cal C} db\frac{b}{2} J_0(b Q_T)e^{S (b,Q)-g_{NP}b^2} 
\left[~\delta H \left(x^0_1,x^0_2;\frac{b_0^2}{b^2}\right)\right.
\label{resum}
\\
&&
+
\left. \frac{\alpha_s (Q^2)}{2\pi} \left\{
\int_{x^0_1}^1  \frac{dz}{z}  \Delta_T C_{qq}^{(1)}(z)
\delta H \left(\frac{x_1^0}{z},x_2^0; \frac{b_0^2}{b^2}\right)
+ \left(x_1^0 \leftrightarrow x_2^0 \right)
\right\} \right],
\non
\eea 
\bea
\lefteqn{\tilde{X}^{\rm NLL} (Q_T^2, Q^2, y)
   = \int_{\cal C} d b \, \frac{b}{2}\,
     J_0 (b Q_T)\, e^{\, S (b , Q)-g_{NP}b^2} 
      \biggl[ H \left(x^0_1,x^0_2;\frac{b_0^2}{b^2}\right)} 
\label{UNPOL}\\
   & &\!\!\!\!\!\!\!\!\!\!\! + \frac{\alpha_s(Q^2)}{2 \pi}
     \left\{ 
          \int_{x_1^0}^1 \frac{dz}{z}  C_{qq}^{(1)} (z)
                    H \left(\frac{x_1^0}{z},x_2^0; \frac{b_0^2}{b^2}\right)
         + \int_{x_1^0}^1 \frac{dz}{z}  C_{qg}^{(1)} (z)
                    K \left(\frac{x_1^0}{z},x_2^0; \frac{b_0^2}{b^2}\right)
+ \left(x_1^0 \leftrightarrow x_2^0 \right)
                   \right\} \biggl] ,
\non
\eea
%
where the DY scaling variables are denoted as
$x_1^0 = \sqrt{Q^2/S}\ e^y$ and $x_2^0 =\sqrt{Q^2/S}\ e^{-y}$, 
$J_0(bQ_T)$ is a Bessel function, 
and $b_0=2e^{-\gamma_E}$ with $\gamma_E$ the Euler constant.  
$\delta H$, $H$, and $K$ denote 
the products of the NLO parton 
distributions
of proton and antiproton, 
summed over the massless quark flavors $q$
with their charge squared $e_q^2$, as 
\be
(\delta) H(x_1,x_2; \mu^2)
= \sum_{q} e_q^2 
\left[(\delta) q(x_1 ,\mu^2) (\delta) q(x_2,\mu^2)
+ (\delta) \bar{q}(x_1 ,\mu^2) (\delta) \bar{q}(x_2, \mu^2)
\right]
\label{tPDF}
\ee 
%
and $K (x_1, x_2; \mu^2) = \sum_q e_q^2\ g(x_1, \mu^2)
\left[ q(x_2, \mu^2) + \bar{q}(x_2, \mu^2) \right]$,
where $\delta q(x, \mu^2 )$, $q(x, \mu^2 )$, and $g(x, \mu^2 )$ 
are
the quark transversity, quark density, and gluon density distributions 
of the proton, respectively; note that 
there is no transversely-polarized gluon distribution at the leading twist.
The Sudakov factor $e^{S(b,Q)}$ 
and the coefficient functions $(\Delta_T) C_{ij}^{(1)}(z)$ 
are perturbatively calculable, and the former gives the all-order resummation of 
logarithmically enhanced contributions
due to multiple emission of soft and/or collinear gluons from the incoming partons.
%
The exponent $S(b,Q)$ is expressed as
%
\begin{equation}
S(b, Q)
=\frac{1}{\alpha_s (Q^2) }h^{(0)}(\lambda)+h^{(1)}(\lambda)\ ,
\label{sudakov:1}
\end{equation} 
where the first and second terms collect the LL and NLL contributions,
respectively, 
in terms of the two functions $h^{(0)}(\lambda)$ and $h^{(1)}(\lambda)$, 
which depend on $\alpha_s(Q^2)$ only through
\begin{equation}
\lambda =  
\beta_0\alpha_s( Q^2 ) \ln(Q^2b^2/b_0^2+1)
\equiv 
\beta_0 \alpha_s( Q^2 ) \tilde{L}\ ,
\label{eq:lambda}
\end{equation} 
with $\beta_0$ the first coefficient of the QCD $\beta$ function.
When $Q_T \ll Q$,  $\tilde{L}$ plays the role of the large-logarithmic
expansion parameter in the $b$ space, as $b\sim 1/Q_T$. 
In this relevant region, 
$\lambda$ can be as large as 1 even for $\alpha_s( Q^2 ) \ll 1$, and
the ratio of two terms in 
(\ref{sudakov:1}) is of ${\cal O}(\alpha_s )$; note that
the NNLL or higher-level corrections, which are down by $\alpha_s$ or more, are neglected in 
(\ref{sudakov:1})~\cite{KKT:07,KKT:07-2}.
The Sudakov exponent is independent of process as well as scheme 
to the NLL accuracy~\cite{KKST:06,BCDeG:03,KT:82}, 
so that (\ref{sudakov:1}) gives the universal Sudakov factor common to
(\ref{resum}) and (\ref{UNPOL}).
%
%
The explicit form of $h^{(0)}(\lambda)$ and $h^{(1)}(\lambda)$,
as well as the coefficient functions
$\left(\Delta_T \right)C_{ij}^{(1)}(z)$
in the $\overline{\rm MS}$ scheme, 
can be found in \cite{KKST:06,KKT:07,KKT:07-2}.
%
It is implicit in (\ref{resum}) and (\ref{UNPOL}) that  
the $b$ dependence of $\delta H (x_1, x_2; b_0^2/b^2)$, 
$H (x_1, x_2; b_0^2/b^2)$ and $K (x_1, x_2; b_0^2/b^2)$,
associated with
the NLO perturbative evolution of the parton distributions 
from the factorization scale $\mu_F=Q$ to the scale $b_0/b$, is
also organized in terms of (\ref{eq:lambda}) to ensure the consistent NLL
accuracy; i.e., the customary NLO evolution operators for the distributions 
are expanded up to the NLL term, 
where the LL term proves to be absent~\cite{KKT:07,KKT:07-2}.
The Fourier transformation to the $Q_T$ space in (\ref{resum}), (\ref{UNPOL})
is performed along a contour ${\cal C}$ 
in the complex $b$ space~\cite{KKT:07,KKT:07-2,LKSV:01,BCDeG:03},
avoiding the singularity of the Sudakov exponent (\ref{sudakov:1})
at $\lambda =1$, which is 
associated with 
the Landau pole in the perturbative running coupling.
This singularity in the $b$ space 
signals the onset of additional nonperturbative phenomena at 
very large values of $|b|$, and the corresponding nonperturbative
effects are complemented in (\ref{resum}), (\ref{UNPOL}) by
introducing a Gaussian smearing function $\exp( -g_{NP} b^2)$
with a parameter $g_{NP}$, following 
the usual procedure~\cite{KKST:06,BCDeG:03,LKSV:01,CSS:85}.
This may be interpreted as representing ``intrinsic transverse momentum'' 
of partons inside proton, and
we use the same smearing
function
for both polarized and unpolarized cases,
following our previous works~\cite{KKT:07,KKT:07-2} for tDY in $pp$ collisions.

The ``regular component'' $(\Delta_T) \tilde{Y}$ 
is determined by the matching procedure 
expanding (\ref{resum}), (\ref{UNPOL})
in powers of $\alpha_s(Q^2 )$ and assuming $g_{NP} \rightarrow 0$ 
in perturbation theory,
so that (\ref{NLL+LO})  
coincides exactly with the fixed-order result of the corresponding 
polarized and unpolarized differential cross sections, 
up to ${\cal O}(\alpha_s)$~\cite{KKST:06,LKSV:01,BCDeG:03}. 
Namely, the LO cross section for $Q_T >0$, 
$(\Delta_T) d \sigma^{\rm LO}/d Q^2 d Q_T^2 d y d \phi$,
which is of ${\cal O}(\alpha_s)$
because the finite $Q_T$ of the lepton pair is provided by the recoil 
from the gluon radiation, is given by (\ref{NLL+LO}) with
the replacement 
$(\Delta_T) \tilde{X}^{\rm NLL} \ra (\Delta_T) \tilde{X}^{\rm NLL} |_{\rm FO}$, 
where $(\Delta_T) \tilde{X}^{\rm NLL} |_{\rm FO}$
denotes the terms resulting from the expansion of the resummed expression up to the
fixed-order $\alpha_s(Q^2 )$. 
According to the structure of (\ref{NLL+LO}) via the matching with the LO cross sections,
we refer to (\ref{NLL+LO}) as the ``NLL+LO'' prediction,
and this gives 
the tDY differential cross sections in the $\overline{\rm MS}$ scheme, which are
well-defined over the entire range of $Q_T$. 


The ratio from (\ref{NLL+LO})
yields the double transverse-spin asymmetry in tDY as
\bea
\aqt=\frac{1}{2} \cos(2\phi) 
\frac{\Delta_T\tilde{X}^{\rm NLL}(Q_T^2 , Q^2, y)
+\Delta_T\tilde{Y}(Q_T^2, Q^2, y)}
{\tilde{X}^{\rm NLL}(Q_T^2, Q^2, y)+\tilde{Y}(Q_T^2, Q^2, y)}\ ,
\label{asym}
\eea
for measured $Q_T$, $Q$, $y$, and $\phi$.
To the fixed-order $\alpha_s$ without the soft gluon resummation, 
$\left(\Delta_T \right) \tilde{X}^{\rm NLL} \rightarrow  
\left(\Delta_T \right) \tilde{X}^{\rm NLL} |_{\rm FO}$ as discussed above,
and (\ref{asym}) reduces to 
the LO prediction $\aqtn^{\rm LO}(Q_T)$
for $Q_T >0$.
We also introduce the asymmetry in terms of the NLL resummed components, 
(\ref{resum}) and (\ref{UNPOL}),
which, respectively, dominate the numerator and denominator 
in (\ref{asym}) when $Q_T \ll Q$:
\begin{equation}
\aqtn^{\rm NLL}(Q_T) = \frac{1}{2}\cos(2\phi)
\frac{\Delta_T \tilde{X}^{\rm NLL} (Q_T^2 , Q^2 ,y)}{\tilde{X}^{\rm NLL} 
(Q_T^2 , Q^2 ,y)}\ .
\label{asymNLL}
\end{equation}
As emphasized in \cite{KKT:07,KKT:07-2} in the context of the $pp$ collisions, 
the $Q_T \rightarrow 0$ limit of (\ref{resum}) and (\ref{UNPOL}) 
deserves special attention:
at $Q_T = 0$, the $b$ integral of (\ref{resum}) and (\ref{UNPOL})
is controlled by a saddle point and can be evaluated analytically as 
\be
(\Delta_T) \tilde{X}^{\rm NLL} (0 , Q^2, y)
=\left[\frac{b_0^2}{4Q^2 \beta_0 \alpha_s(Q^2)} 
\sqrt{\frac{2\pi}{{\zeta^{(0)}}''(\lambda_{SP})}} 
e^{-\zeta^{(0)}(\lambda_{SP})+h^{(1)}(\lambda_{SP})} \right] 
(\delta) H \left(x_1^0,x_2^0; \frac{b_0^2}{b_{SP}^2}\right) ,
\label{eq:speval}
\ee
where 
$\zeta^{(0)}(\lambda) \equiv - \lambda/[\beta_0 \alpha_s(Q^2)]
-h^{(0)}(\lambda)/ \alpha_s (Q^2)
+ [g_{NP}b_0^2 /Q^2 ]e^{\lambda/[ \beta_0 \alpha_s (Q^2 )]}$,
and the saddle-point value
of (\ref{eq:lambda})\footnote{
For the kinematics of our interest, we have the saddle point well above $b=0$, 
and we can use the definition 
$\lambda = \beta_0\alpha_s( Q^2 ) \ln(Q^2b^2/b_0^2)$,
up to the exponentially suppressed corrections
to (\ref{eq:speval}) (see \cite{KKT:07,KKT:07-2}).}, 
$\lambda_{SP}= \beta_0\alpha_s( Q^2 ) \ln(Q^2b_{SP}^2/b_0^2 )$,
is defined by the condition ${\zeta^{(0)}}' (\lambda_{SP} )=0$.
The saddle-point formula (\ref{eq:speval}) 
is exact up to the ${\cal O}(\alpha_s)$ corrections that actually correspond 
to the NNLL contributions in the region $Q_T \approx 0$ 
(see the discussion below (\ref{eq:lambda}), and also 
\cite{KKT:07,KKT:07-2} for the details). 
Note that the gluon distribution completely decouples from 
$\tilde{X}^{\rm NLL} (0 , Q^2, y)$ at the NLL accuracy for $Q_T \approx 0$.
The prefactor inside the square bracket in the RHS
involves ``large perturbative effects'' due to the Sudakov factor, as well as
the Gaussian smearing factor with $g_{NP}$;
the former contribution drives the well-known asymptotic behavior~\cite{PP} 
of the DY cross sections, $\sim (\Lambda_{\rm QCD}^2 /Q^2 )^{a\ln(1+1/a)}$ 
with $a\equiv A_q^{(1)}/(2\pi \beta_0)$, for $Q \gg \Lambda_{\rm QCD}$.
Because this prefactor is common to both the polarized and unpolarized cross sections, 
we obtain the remarkably compact formula for the $Q_T \rightarrow 0$ limit 
of (\ref{asymNLL})~\cite{KKT:07,KKT:07-2}:
\begin{equation}
\aqtn^{\rm NLL}(Q_T =0)=\frac{1}{2}\cos(2 \phi ) 
\frac{\delta H \left( x_1^0, x_2^0;\ b_0^2 /b_{SP}^2 \right)}
{H \left( x_1^0, x_2^0;\ b_0^2 /b_{SP}^2 \right)}\ .
\label{eq:attnll}
\end{equation}

Using the 
formulae described above, we study the behavior of tDY
to be observed in $p\bar{p}$ collisions at GSI. 
The PAX Collaboration has proposed 
the tDY experiments
in $p\bar{p}$ collisions at $S=30$ and $45$~GeV$^2$ in the fixed-target mode, 
and those up to $S= 210$~GeV$^2$ in the collider mode~\cite{PAX:05}.
Those GSI-PAX experiments will probe $0.2 \lesssim Q/\sqrt{S}\lesssim 0.7$, and thus 
the transversities in the ``valence region'' 
in a wide range of $x$. 
To compute 
the tDY cross sections (\ref{NLL+LO}) at the NLL+LO accuracy with these GSI kinematics, 
we have to specify the NLO parton distributions to be substituted. 
We use the NLO GRV98 distributions~\cite{GRV:98} for the unpolarized quark and 
gluon distributions $q(x,\mu^2)$ and $g(x,\mu^2)$.
For the NLO transversity distributions $\delta q(x,\mu^2)$,
we consider the two typical assumptions that have been used 
in the literature~\cite{BDR:02,Anselmino:2004ki,MSSV:98,KKST:06,
KKT:07,KKT:07-2,BCCGR:06,SSVY:05}:
at a low input scale $\mu_0 \lesssim 1$~GeV, these assume the saturation of 
Soffer's inequality~\cite{Soffer:95} 
as $\delta q(x,\mu_0^2)=[q(x,\mu_0^2)+\Delta q(x,\mu_0^2)]/2$,
and the relation,
%
\be
\delta q(x,\mu_0^2)=\Delta q(x,\mu_0^2)\ ,
\label{input}
\ee
exact in the non-relativistic limit, respectively, 
and their QCD evolution from $\mu_0$ to a higher scale $\mu$ is controlled by 
the NLO DGLAP kernel~\cite{KMHKKV:97} for the transversity;
here $\Delta q(x,\mu_0^2)$ denote the longitudinally polarized quark distributions.
The first case yields $\delta q(x,\mu^2 )$
that satisfies
Soffer's inequality\footnote{
The obtained $\delta q(x,\mu^2 )$ is actually very close to 
$[q(x,\mu^2)+\Delta q(x,\mu^2)]/2$,
except for small $x$ ($\lesssim 0.2$).} 
and provides an upper bound on the transversities~\cite{MSSV:98}.
The second case (\ref{input}) is suggested also by the estimates 
from relativistic quark models for nucleon~\cite{BDR:02,Anselmino:2004ki,waka}.
For the input functions in the RHS of these two assumptions,
we take the NLO GRV98 distributions $q(x,\mu_0^2)$, as noted above, 
and GRSV2000 (``standard scenario'') distributions 
$\Delta q(x,\mu_0^2)$ \cite{GRSV:00} with $\mu_0^2 = 0.40$ GeV$^2$.
%
\begin{figure}
\bc
\includegraphics[height=5.8cm]{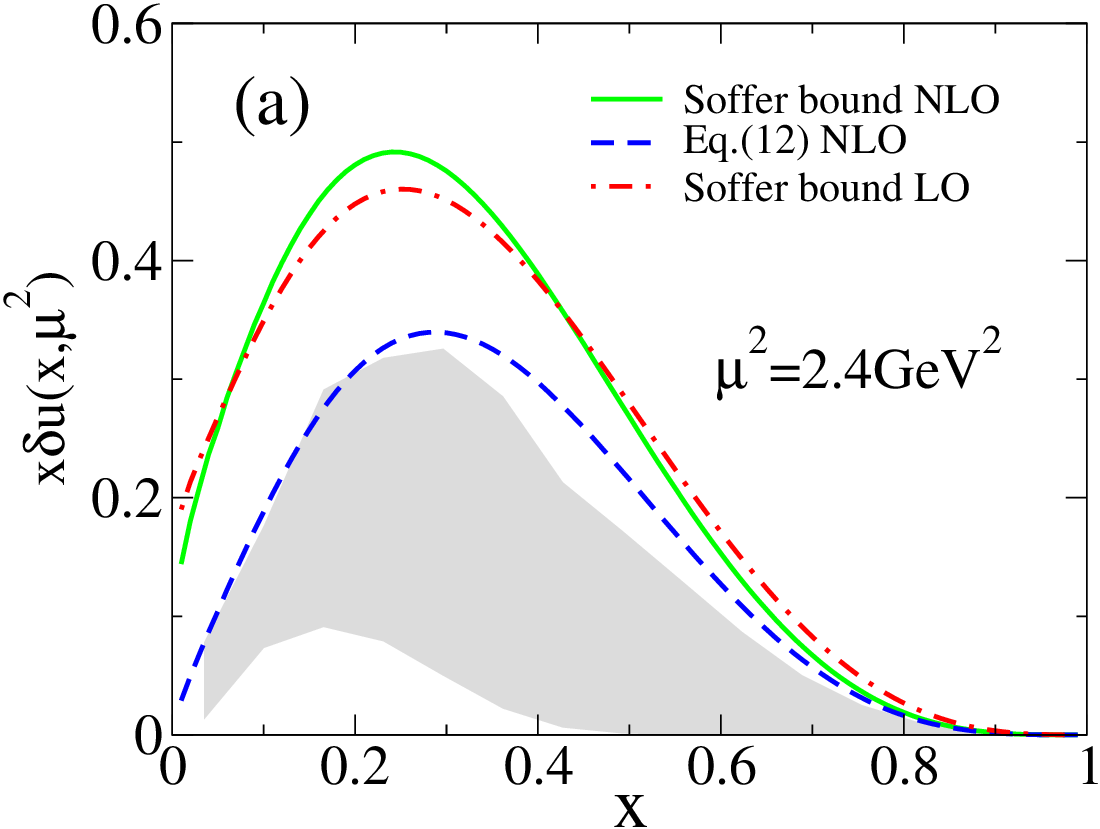}~~~~
\includegraphics[height=5.8cm]{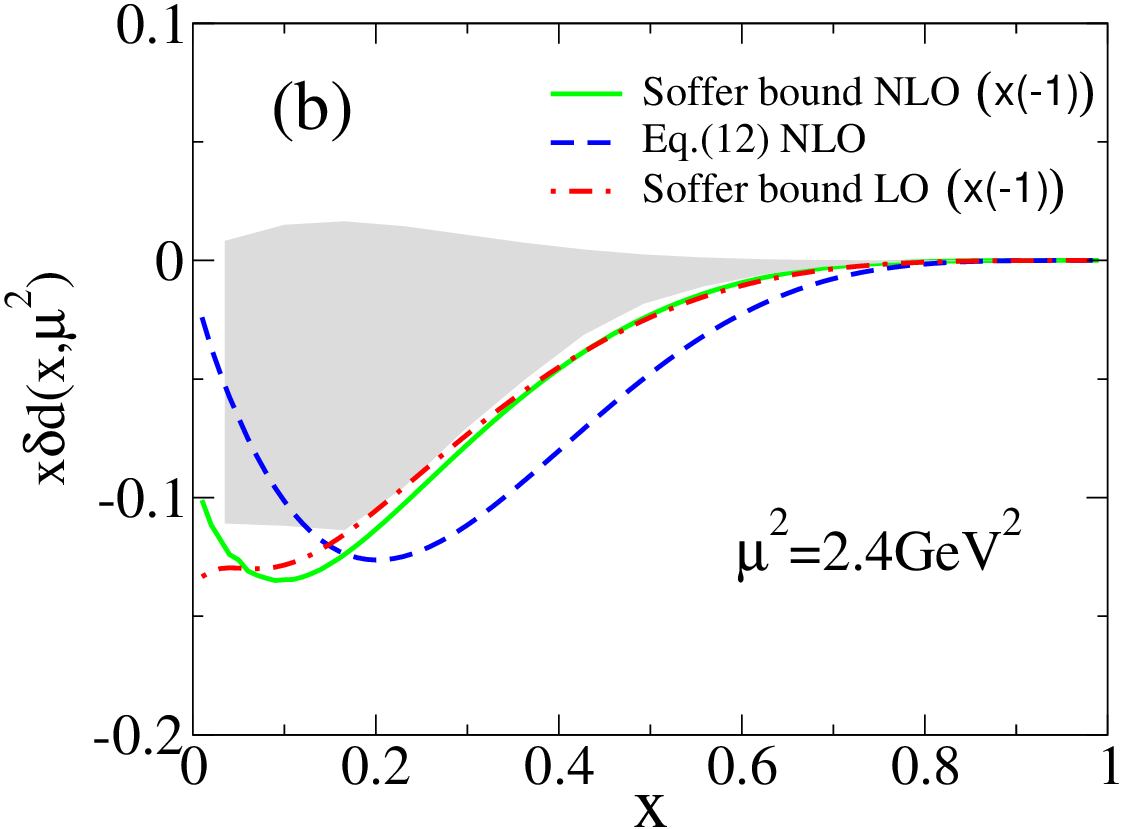}
\ec
\caption{The transversity distributions for (a) $u$-quark and (b) $d$-quark.
The solid and dashed curves plot the NLO distributions 
corresponding to the Soffer bound and the relation (\ref{input}), 
respectively.
The dot-dashed curve shows the LO distribution corresponding to the Soffer bound 
and the shaded area shows the result of the LO global fit in \cite{Anselmino:07}.}
\label{fig:1}
\end{figure}
%
The obtained NLO transversity distributions for $u$ and $d$ quarks, 
$x\delta u(x,\mu^2 )$ and $x\delta d(x,\mu^2 )$,
are shown in Figs.~1~(a) and (b), respectively, as a function of $x$
with $\mu^2 =2.4$ GeV$^2$. The solid curve shows the result corresponding 
to the Soffer bound,
%
and the dashed curve shows the result using (\ref{input}).
For comparison, we also depict the shaded area which represents the result 
(a one-sigma confidence interval) for the LO transversity distributions 
extracted through the global fit to the data~\cite{Anselmino:07}, and
the dot-dashed curve which shows the LO transversity distributions 
corresponding to the Soffer bound with the LO inputs of \cite{GRV:98,GRSV:00}.
We note that the above assumption for the Soffer bound, 
$\delta q(x,\mu_0^2)=[q(x,\mu_0^2)+\Delta q(x,\mu_0^2)]/2$,
following the literature~\cite{MSSV:98,KKST:06,KKT:07,KKT:07-2,BCCGR:06,SSVY:05}
yields the positive polarization for the $d$ quark, 
$\delta d(x,\mu^2 )>0$, in contrast to the result using (\ref{input}),
but the sign of the polarization cannot be detected in tDY for
$p\bar{p}$ collision (see~(\ref{tPDF})).
For convenience of comparison, the solid and dot-dashed curves in Fig.~1(b) 
show the results multiplied by $-1$.

The results in Figs.~1~(a) and (b) indicate that the empirical LO transversities 
are smaller compared with the transversities corresponding to 
the Soffer bound, in particular for the $u$ quark.
On the other hand, the NLO transversities using (\ref{input}) 
lie slightly outside the one-sigma error bounds of the empirical fit.
The two NLO sets as well as the empirical fit have 
$\left(\delta u(x,\mu^2 )\right)^2 \gg \left( \delta d(x,\mu^2 )\right)^2$
in the valence region relevant at GSI kinematics, 
so that
\be
\delta H(x_1,x_2;\mu^2) \simeq e_u^2 
\delta u(x_1 ,\mu^2) \delta u(x_2,\mu^2)\ ,
\label{valence}
\ee
for (\ref{tPDF}).
Hence the cross section (\ref{NLL+LO}) and the asymmetry (\ref{asym}) at GSI
allow a direct access to $|\delta u(x,\mu^2 )|$.
The NLO transversities using (\ref{input}) satisfy Soffer's inequality 
for the $u$ quark, but violate it for the $d$ quark by a small amount, 
see the last footnote and Figs.~1~(a), (b)\footnote{
The corresponding $\bar{u}$-, $\bar{d}$-, and $\bar{s}$-quark distributions 
satisfy Soffer's inequality.}.
However, this violation of Soffer's inequality will be harmless to our
numerical estimates of the cross sections and asymmetries because of 
the dominance of the $u$-quark distribution noted above. 
We also mention that, strictly speaking, 
Soffer's inequality for the NLO distributions receives
the additional scheme-dependent radiative corrections~\cite{Soffer:95},
and the corresponding corrections would modify the solid curve in Figs.~1~(a), 
(b) by a certain amount of ${\cal O}(\alpha_s)$. 

\begin{figure}[!t]
\bc
\includegraphics[height=5.8cm]{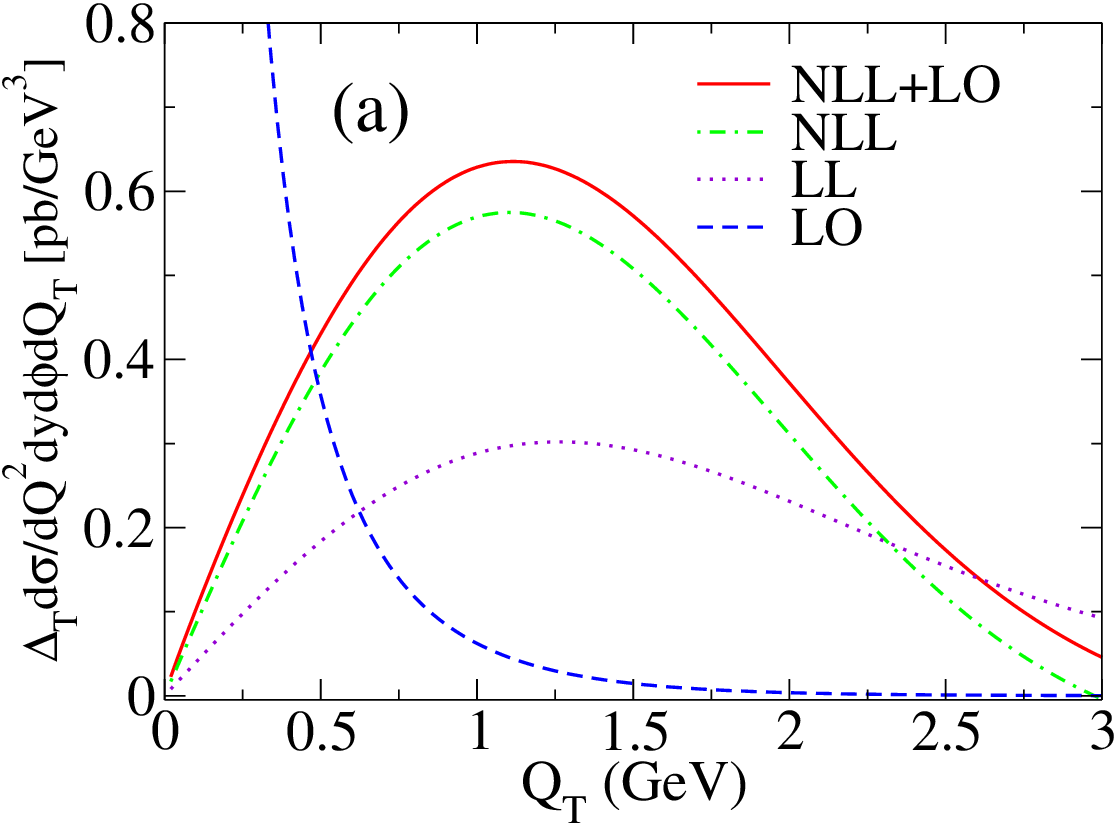}~~~~
\includegraphics[height=5.8cm]{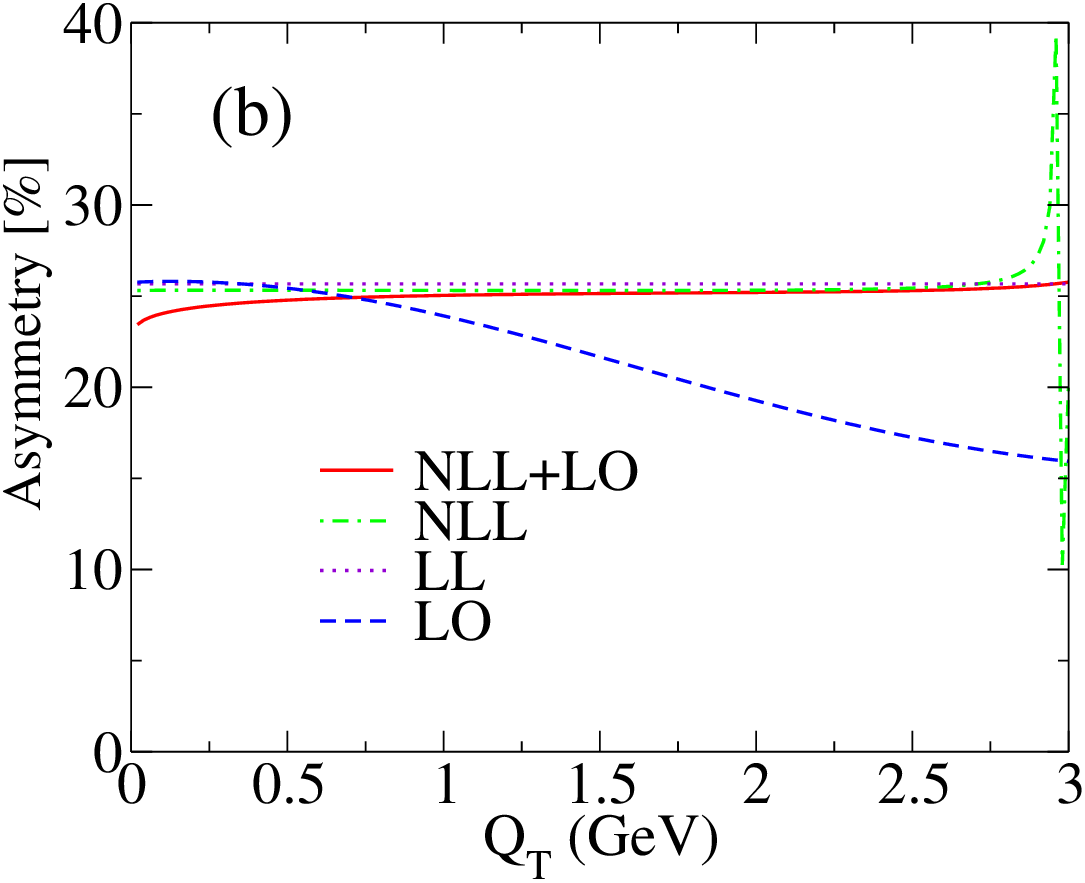}
\ec
\caption{The tDY with GSI kinematics, $S=210$ GeV$^2$, $Q=4$ GeV,
$y=0$ and $\phi=0$, and with $g_{NP}=0.5$ GeV$^2$, using the NLO 
transversity distributions which correspond to the Soffer bound: 
(a) The spin-dependent part of the cross section, $\Delta_Td\sigma/dQ^2dQ_Tdyd\phi$.
(b) The double transverse-spin asymmetries obtained using each curve in (a).
}
\label{fig:2}
\end{figure}

In all the following numerical evaluation, we choose $\phi=0$ for the
azimuthal angle of a lepton; extension to other $\phi$ is straightforward by 
taking into account 
the $\cos(2\phi)$ dependence displayed  
in the relevant formulae (\ref{NLL+LO}), (\ref{asym}), etc.
For the nonperturbative parameter in (\ref{resum}) and (\ref{UNPOL}), 
we use $g_{NP}=0.5$ GeV$^2$ 
as used for the RHIC and J-PARC cases~\cite{KKST:06,KKT:07,KKT:07-2}.

Probing the transversity distributions through the tDY 
in the collider mode is more promising than that in the fixed-target mode, because 
at higher energies 
the description 
based on QCD perturbation theory is supposed to be more accurate~\cite{SSVY:05}.   
In Fig.~2(a), we show the $Q_T$ spectrum of the dilepton in tDY 
with $S=210$ GeV$^2$, $Q=4$ GeV, and $y=0$, corresponding to a
kinematics in the collider mode.
We use the transversity distributions corresponding to
the Soffer bound (the solid curve in Fig.~1),
which were used in similar study of the $Q_T$ spectrum 
at RHIC~\cite{KKST:06,KKT:07,KKT:07-2}.
The solid line shows the spin-dependent part of the NLL+LO differential 
cross section (\ref{NLL+LO}), multiplied by $2Q_T$, and 
the dot-dashed line plots the contribution from the NLL resummed part 
$\Delta_T\tilde{X}^{\rm NLL}$ given in (\ref{resum}).  
The dotted line plots the contribution from the LL resummed part, 
\be
\Delta_T \tilde{X}^{\rm LL} (Q_T^2, Q^2, y)
=
\left[\int_{\cal C} db\frac{b}{2} J_0(b Q_T)
e^{h^{(0)}(\lambda)/\alpha_s (Q^2) -g_{NP}b^2} \right]
\delta H \left(x^0_1,x^0_2; Q^2 \right)\ ,
\label{resumLL}
\ee
which is obtained from (\ref{resum}) by omitting the NLL terms 
with the nonperturbative inputs ($g_{NP}$ and transversity distributions) kept intact;
note that the $b$ dependence of the parton distributions in (\ref{resum})
is associated with the NLL-level terms, as mentioned below 
(\ref{eq:lambda})~\cite{KKT:07,KKT:07-2,BCDeG:03}.
The dashed line shows the $Q_T$ spectrum using the LO cross section
$\Delta_T d \sigma^{\rm LO}/d Q^2 d Q_T^2 d y d \phi$ introduced above (\ref{asym}), 
which is divergent as $Q_T\rightarrow 0$ due to the singular terms 
$\propto \ln(Q^2/Q_T^2)/Q_T^2$, $\propto 1/Q_T^2$.
By resumming the singular large logarithms to all orders in $\alpha_s$,
the $Q_T$ spectra are completely redistributed and well behaved,
forming a peak at $Q_T\sim 1$~GeV.
In fact, around the peak region,
the NLL+LO cross section is dominated by the contribution from the NLL
resummed part $\Delta_T\tilde{X}^{\rm NLL}$.
%
%
It is also remarkable that the NLL result is considerably enhanced
compared with  
the LL result, though the integrations of these two results over $Q_T$ coincide,
using 
$\lambda = 0$ at $b=0$ (see (\ref{eq:lambda}) and \cite{KKT:07,KKT:07-2}), 
up to
the ${\cal O}\left(\alpha_s(Q^2)\right)$ corrections
associated with 
the coefficient function $\Delta_T C^{(1)}_{qq}$.
We note that the pattern similar to Fig.~2(a) is observed also in the 
corresponding $Q_T$ spectra for the unpolarized differential cross sections.

The behavior of the NLL+LO cross section 
at $Q_T\simeq 3$ GeV in Fig.~2(a) 
is still affected by the ``broadening'' due to the nonperturbative smearing 
$e^{-g_{NP}b^2}$ in (\ref{resum});
as a result, 
the NLL+LO cross section is larger than
the LO one at $Q_T\simeq 3$ GeV, although the former was matched to the latter 
in the region where the logarithm $\ln(Q^2/Q_T^2)$ is not large
(see the discussion above (\ref{asym})).
Because the separation between the ``peak region'' $Q_T \sim 1$~GeV
and the ``matching region'' $Q_T \sim Q$ is not so large for the present kinematics,
the value $Q_T\simeq 3$~GeV actually corresponds to the boundary 
between ``smeared''
and purely perturbative 
regimes.
In fact, for higher $\sqrt{S}$ and $Q$ 
such that the peak and matching regions are largely separated,
the NLL+LO cross section around $Q_T\simeq Q$ is independent of 
the nonperturbative smearing 
and reduces to the LO one (see e.g. the results for RHIC in 
\cite{KKST:06,KKT:07,KKT:07-2}).
Another consequence due to the moderate-energy kinematics
is that
the perturbative contributions of ${\cal O}(\alpha_s^2)$ and higher, involved in the 
NLL resummed component~(\ref{resum}), are not negligible around $Q_T\simeq Q$
at the quantitative level, as $\alpha_s(Q^2)/\pi \sim 0.07$.
It is worth noting that, even in 
much higher energy processes such as 
$Z$ boson~\cite{LKSV:01} and Higgs boson productions~\cite{BCDeG:03},
the higher-order
perturbative contributions in the resummed component remain numerically sizable
also at rather large $Q_T$ well above the peak region. 
In the present case, however, the impact of those higher-order
contributions can be even more significant, because 
the LO cross section at $Q_T\sim Q$ is very small (see Fig.~2(a));
it is actually much smaller
than its canonical size,
$[\alpha^2 /3 N_c S Q^2]\times [\alpha_s (Q^2)/\pi Q] 
\sim 10^{-2}$ pb/GeV$^3$ (see (\ref{NLL+LO})),
by the extra factor $\sim 0.01$; 
here the extra small factor comes from the fact that,
for the moderate energy $\sqrt{S}=14.5$ GeV,
the $2\rightarrow 3$ partonic processes emitting a real gluon  
with the ``high'' transverse
momentum $k_T \sim Q$ are possible only from the initial-state partons 
with the large momentum fractions $x_{1,2}$ as $x_1 x_2  \gtrsim 0.3$ ($\gg Q^2/S$),
for which the products of the parton distributions in (\ref{tPDF}) 
are strongly suppressed (see Fig.~1).
These points suggest that the accurate quantitative description 
around 
$Q_T\simeq Q$ at GSI would eventually require the matching procedure
taking into account also the ${\cal O}(\alpha_s^2)$
or higher contributions 
in perturbation theory,
which is beyond our NLL+LO framework. 
One may expect that the mechanisms observed 
for the large $Q_T$ region in the ``NLL+NLO'' cross section of $Z$ 
boson production~\cite{LKSV:01}
and in the ``NNLL+NLO'' cross section of Higgs boson production~\cite{BCDeG:03} 
could also provide a better treatment for the matching region $Q_T\sim Q$ 
in the present case.
Nevertheless, in this paper, we insist on the NLL+LO framework
because the tDY $Q_T$-differential cross section is not known at NLO,
and we restrict our quantitative discussion below the matching region.
%

Figure~2(b) shows the asymmetries as functions of $Q_T$,
obtained by taking the ratio of each curve in Fig.2(a) to 
the corresponding $Q_T$ spectra for the unpolarized cross sections; i.e., 
the solid, dot-dashed, 
and dashed curves plot the NLL+LO ($\aqt$ of (\ref{asym})),
NLL ($\aqtn^{\rm NLL}(Q_T)$ of (\ref{asymNLL})),
and LO ($\aqtn^{\rm LO}(Q_T)$ below (\ref{asym})) asymmetries, respectively,
and the dotted curve shows 
the LL asymmetry, obtained by using (\ref{resumLL}) and the similar 
formula for $\tilde{X}^{\rm LL}$, as
\be
\aqtn^{\rm LL}
= \frac{1}{2}\cos(2\phi)
\frac{\Delta_T \tilde{X}^{\rm LL} (Q_T^2 , Q^2 ,y)}{\tilde{X}^{\rm LL} 
(Q_T^2 , Q^2 ,y)} 
=\frac{1}{2}\cos(2\phi)
\frac{\delta H(x_1^0,x_2^0; Q^2)}{H(x_1^0,x_2^0; Q^2)}\ ,
\label{asymLL}
\ee
which is constant in $Q_T$.
All asymmetries in Fig.~2(b), except the LO result, show similar flat behavior
with almost the same value $\simeq25$\%. 
The results observed in Fig.~2(a) imply that these flat behaviors 
are governed by the soft gluon resummation contributions.
In particular, we have, for $Q_T$ around the peak region in Fig.~2(a),
\be
\aqt \simeq \aqtn^{\rm NLL}(Q_T) \simeq \aqtn^{\rm NLL}(0)  \ ,
\label{fullNLL}
\ee
and the flat behavior, represented by the second (approximate) equality,
reflects the fact that the soft gluon effects resummed into the Sudakov factor 
$e^{S(b,Q)}$ 
are universal to the NLL accuracy
between $\Delta_T\tilde{X}^{\rm NLL}$ and $\tilde{X}^{\rm NLL}$ in (\ref{asymNLL}). 
%
In fact, 
similar 
flat behavior 
is observed 
in $pp$ collisions at RHIC and J-PARC kinematics~\cite{KKT:07,KKT:07-2};
note, remarkably, the present result for $p\bar{p}$ collisions turns out to be
even flatter.
%
As $Q_T\rightarrow 0$, away from the peak region of the cross section, 
$\aqt$ decreases slightly due to 
the terms $\propto\ln(Q^2/Q_T^2)/Q^2$ contained in 
$\Delta_T \tilde{Y}$ and $\tilde{Y}$ in (\ref{asym}).
The difference between $\aqt$ and the LO asymmetry at $Q_T\simeq 3$~GeV reflects 
the above-mentioned discrepancy between the corresponding cross sections. 

We here recall that, in the $pp$-collision cases at RHIC and J-PARC
using the same input transversity distributions as in Fig.~2(b),
the significant enhancement of $\aqtn^{\rm NLL}(Q_T)$ ($\simeq \aqt$) 
compared with $\aqtn^{\rm LL}$ has been found~\cite{KKT:07,KKT:07-2}.
Such enhancement
is not seen in the present $p\bar{p}$-collision case.   
This fact indicates that the large NLL-level corrections 
shown in Fig.~2(a) are canceled in the asymmetry $\aqtn^{\rm NLL}(Q_T)$ with the
corresponding corrections to the unpolarized cross section,
although this is not the case in the $pp$ collisions.


%
\begin{figure}
\bc
\includegraphics[height=5.8cm]{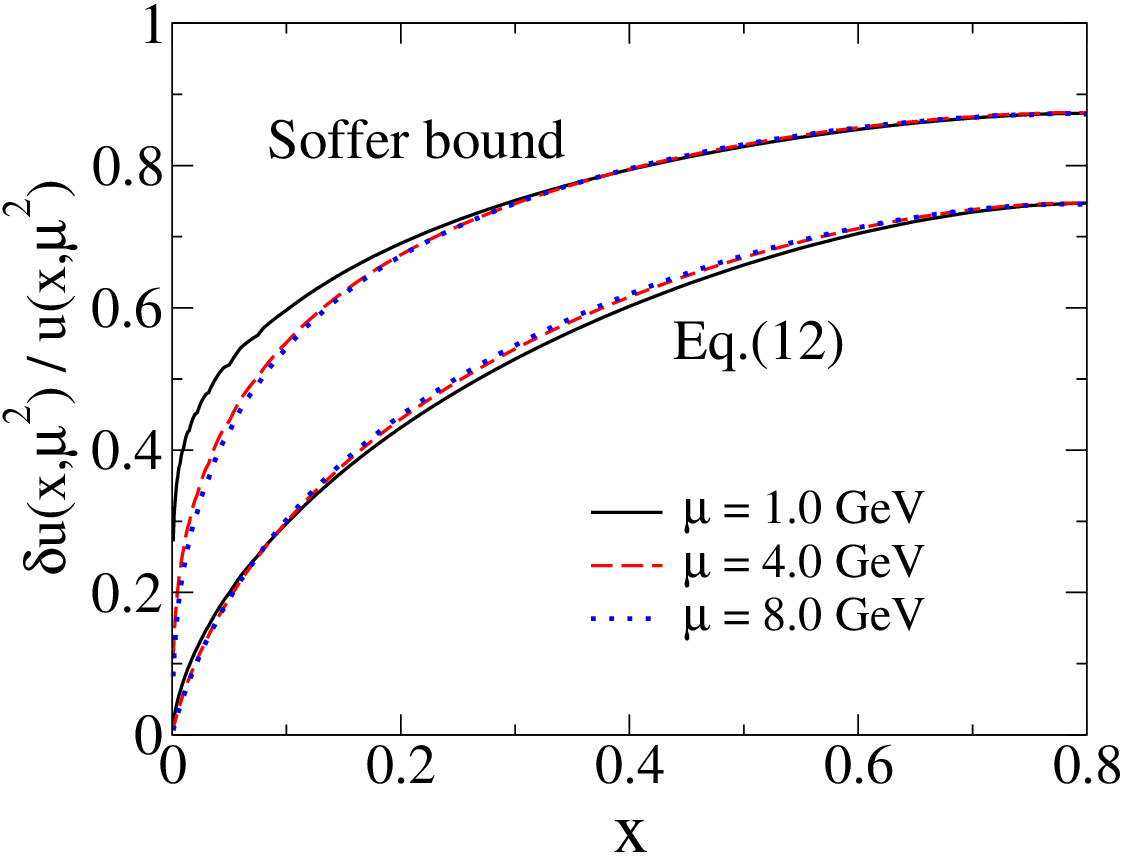}
\ec
\caption{The ratio of the transversity to unpolarized quark
distribution for $u$-quark at different scales $\mu$. The upper and lower curves are
obtained using the NLO transversity distributions corresponding to 
the Soffer bound and the relation (\ref{input}), respectively.
}
\label{fig:3}
\end{figure}
%
To clarify the reason behind this remarkable difference between the
$p\bar{p}$- and $pp$-collision cases, the saddle-point formula 
(\ref{eq:attnll}) is useful.
Similarly to the $pp$-collision case~\cite{KKT:07,KKT:07-2}, 
the property (\ref{fullNLL}) allows us to use (\ref{eq:attnll}) as a
sufficiently accurate estimation of
$\aqt$, $\aqtn^{\rm NLL}(Q_T)$ (see Fig.~4 and Table~1 below).
We obtain $b_0/b_{SP} =1.0$ GeV for the scale of the transversity and unpolarized
distributions in the numerator and denominator of (\ref{eq:attnll}) at the kinematics 
of Fig.~2, using the condition 
that was noted below (\ref{eq:speval})\footnote{
The value of $b_0/b_{SP}$ in principle depends on the input values for $Q$ and
$g_{NP}$, but in practice $b_0/b_{SP}\simeq 1$ GeV,
irrespective of $Q$ and $g_{NP}$ (see \cite{KKT:07,KKT:07-2} for the detail).
This in turn implies that (\ref{eq:attnll}), and thus $\aqt$ as well as 
$\aqtn^{\rm NLL}(Q_T)$, are almost independent of the value of $g_{NP}$. 
We have explicitly checked this point also in the direct numerical evaluation
of (\ref{asym}) with (\ref{resum}), (\ref{UNPOL})
in the range $g_{NP}=0.3$-$0.8$ GeV$^2$.}.
%
$\aqtn^{\rm LL}$ of (\ref{asymLL}) is different from (\ref{eq:attnll}),
only in the scale $Q$ ($=4$ GeV for Fig.~2) of the parton distributions.
Therefore, the possible enhancement of $\aqt$, $\aqtn^{\rm NLL}(Q_T)$  
in comparison with $\aqtn^{\rm LL}$
can be understood as a result of QCD evolution of the parton distributions 
from $b_0/b_{SP}$ to $Q$, with $Q^2 \gg b_0^2/b_{SP}^2$:
in fact, the corresponding enhancement in $pp$ collisions at RHIC and J-PARC 
arises because of very different behavior of the sea-quark components under the evolution
between transversity and unpolarized distributions~\cite{KKT:07,KKT:07-2}; 
note that the transversity distributions obey a non-singlet-type evolution
even for the sea-quark components because there is no gluon transversity distribution.
%
On the other hand, in $p\bar{p}$ collisions at GSI, associated with 
the region $0.2 \lesssim x_{1,2}^0 \lesssim 0.7$, 
the relevant formulae (\ref{eq:attnll}), (\ref{asymLL}) for the asymmetries
are dominated by the contributions from the valence components,
such that 
%
\be 
\aqt \simeq \aqtn^{\rm NLL}(Q_T) 
\simeq\frac{1}{2} \cos(2\phi)
\frac{\delta u(x_1^0,b_0^2/b_{SP}^2) \delta u(x_2^0, b_0^2/b_{SP}^2)}
{u(x_1^0 , b_0^2/b_{SP}^2) u(x_2^0, b_0^2/b_{SP}^2)}\ ,
\label{valence2}
\ee
and 
$\aqtn^{\rm LL}\simeq[\cos(2\phi)/2][\delta u(x_1^0 , Q^2) 
\delta u(x_2^0, Q^2)/u(x_1^0 , Q^2) u(x_2^0, Q^2)]$,
using (\ref{valence}) and the similar relation for the 
unpolarized distributions (see (\ref{tPDF})).
As demonstrated in Fig.~3, the scale dependence of the $u$-quark distributions
cancels between the numerator and denominator 
of (\ref{valence2})
in the relevant ``valence region''. 
This fact
explains why
$\aqt \simeq \aqtn^{\rm NLL}(Q_T) \simeq \aqtn^{\rm LL}$ 
in Fig.~2(b). 
We note that actually the same mechanism arises for all kinematics at GSI, and 
also for the input transversity distributions using (\ref{input}) (see Fig.~3). 
This implies that the property,
$\aqt \simeq \aqtn^{\rm NLL}(Q_T) \simeq \aqtn^{\rm LL}$,
is characteristic of all $p\bar{p}$ collisions at GSI.
Moreover, a similar logic applied to (\ref{resum}) allows us to derive
the second approximate equality of (\ref{fullNLL}): using (\ref{valence}) 
and 
the property 
$\delta u(x_{1,2}^0,b_0^2/b^2)\simeq u(x_{1,2}^0,b_0^2/b^2)
\delta u(x_{1,2}^0,b_0^2/b^2_{SP})/u(x_{1,2}^0,b_0^2/b^2_{SP})$ 
implied by Fig.~3, (\ref{resum}) gives
$\Delta_T \tilde{X}^{\rm NLL} (Q_T^2, Q^2, y)$
$\simeq[ \delta u(x_1^0,b_0^2/b_{SP}^2) \delta u(x_2^0, b_0^2/b_{SP}^2)/
u(x_1^0 , b_0^2/b_{SP}^2) u(x_2^0, b_0^2/b_{SP}^2)]\tilde{X}^{\rm NLL} (Q_T^2, Q^2, y)$,
up to the ${\cal O}(\alpha_s)$ corrections associated with the coefficient functions
$\left(\Delta_T \right)C_{ij}^{(1)}(z)$ in (\ref{resum}) and (\ref{UNPOL}).
This immediately gives the second relation in (\ref{fullNLL}), 
combined with (\ref{asymNLL}) and (\ref{eq:attnll}).
This derivation using the properties in the valence region
also explains why $\aqt$, $\aqtn^{\rm NLL}(Q_T)$ 
in $p\bar{p}$ collisions at GSI
are flatter 
than in $pp$ collisions as noted below (\ref{fullNLL})
(see also Figs.~4-7 below).      
%

\begin{figure}[!t]
\bc
\includegraphics[height=5.8cm]{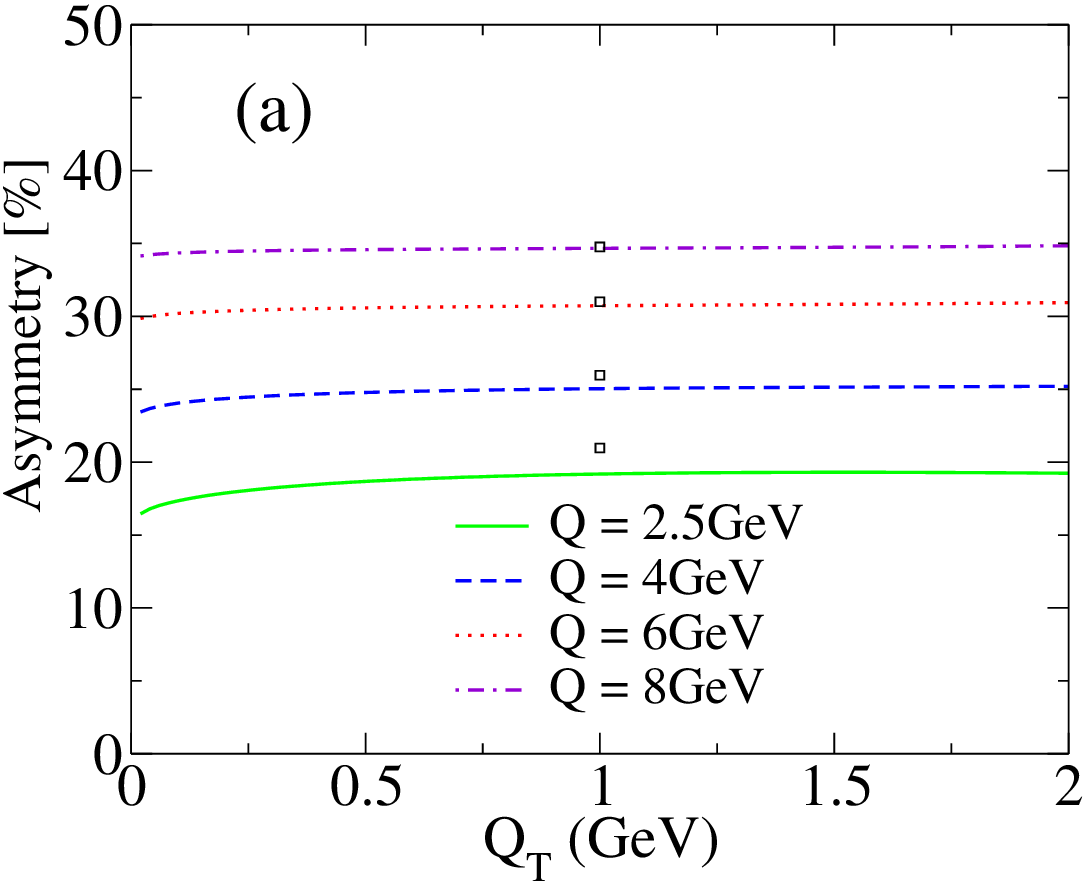}~~~~
\includegraphics[height=5.8cm]{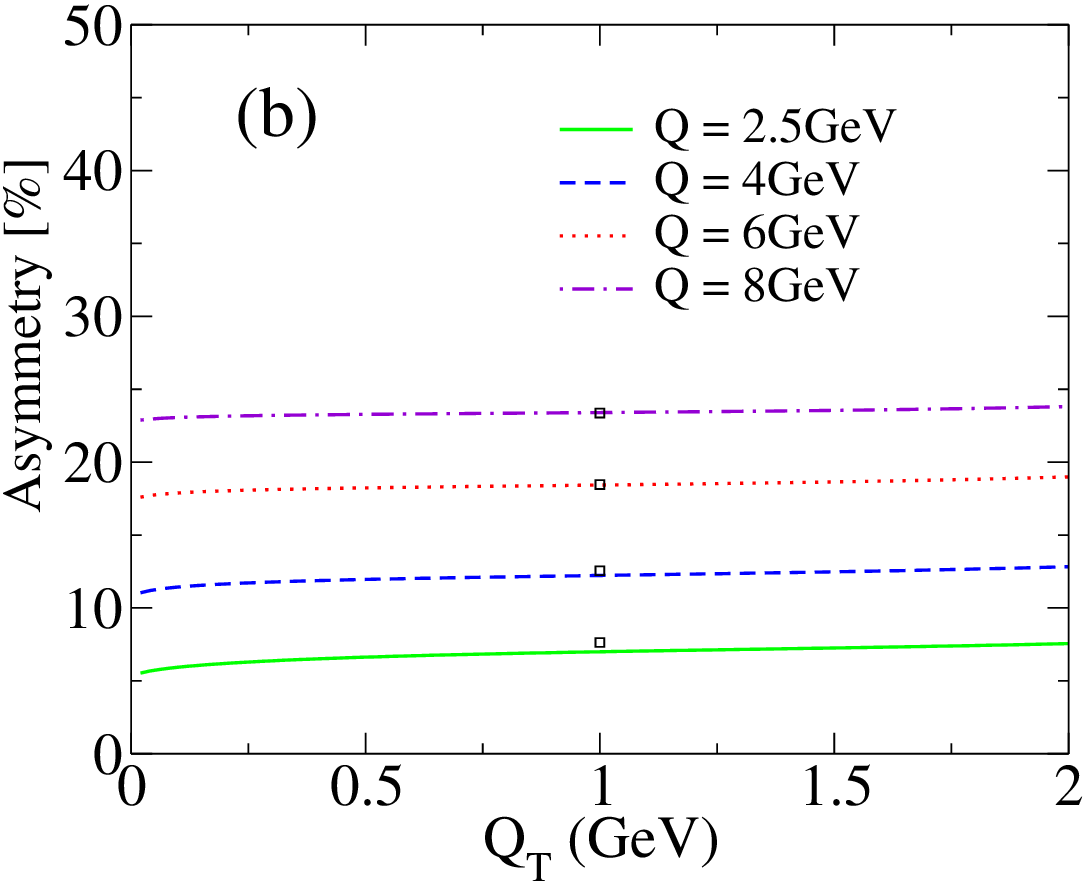}
\ec
\caption{The NLL+LO asymmetry $\aqtn(Q_T)$ of (\ref{asym}) with GSI kinematics,
$S=210$ GeV$^2$ and $y=\phi=0$, and 
with $g_{NP}=0.5$GeV$^2$, using the NLO transversity distributions 
corresponding to (a) the Soffer bound and (b) the relation (\ref{input}), respectively. 
The corresponding values of the saddle-point formula (\ref{eq:attnll}) 
are also shown by the symbols at $Q_T =1$ GeV. 
}
\label{fig:4}
\end{figure}
\begin{table}
[htpb]
\begin{center}
\begin{tabular}{|c|r|r|r|r||r|r|r|r|}
\hline
&\multicolumn{4}{|c||}{Soffer bound}
&
\multicolumn{4}{|c|}{Eq.~(\ref{input})}\\
\hline
$Q$  & 2.5GeV  &  ~4GeV  &  ~6GeV & ~8GeV & 2.5GeV & ~4GeV & ~6GeV & ~8GeV\\
\hline
SP
& 21.0\% & 26.0\% & 31.0\% & 34.8\% & 7.6\% & 12.6\% & 18.5\% & 23.4\%\\
NB
& 19.6\% & 25.3\% & 30.8\% & 34.7\% & 7.2\% & 12.4\% & 18.5\% & 23.4\%\\
\hline
\end{tabular}
\caption{The values of $\aqtn^{\rm NLL}(Q_T=0)$ of (\ref{asymNLL}) for the cases of 
Figs.~\ref{fig:4}~(a) and (b), as ``Soffer bound'' and ``Eq.~(\ref{input})'',
respectively.
SP is obtained using the saddle-point formula 
(\ref{eq:attnll}) and NB is obtained using the numerical $b$-integration
of (\ref{resum}) and (\ref{UNPOL}).}
\label{tab:3}
\end{center}
\end{table}

Using the same nonperturbative and kinematical inputs as in Fig.~2, 
Fig.~4(a) shows the dependence of the NLL+LO asymmetry $\aqt$ of (\ref{asym}) 
on the dilepton mass $Q$;
the dashed curve in Fig.~4(a) is the same as the solid curve in Fig.~2(b), and
the results in Fig.~4(a) 
show the ``maximally possible'' asymmetry, i.e., 
should be considered as optimistic 
estimate (see the solid curve in Fig.~1).
We expect that a more realistic estimate is provided by the results in Fig.~4(b),
which is same as Fig.~4(a) 
but for the case with the input transversity distributions
using (\ref{input}) corresponding to the dashed curve in Fig.~1.
As $Q$ increases, $\aqt$ increase, preserving
the characteristic flat behavior as functions of $Q_T$.
The results 
in Fig.~4(b) using (\ref{input}) are smaller 
compared with the corresponding Soffer bound results in Fig.~4(a),
but still yield rather large asymmetries.

\begin{figure}[t!]
\bc
\includegraphics[height=5.8cm]{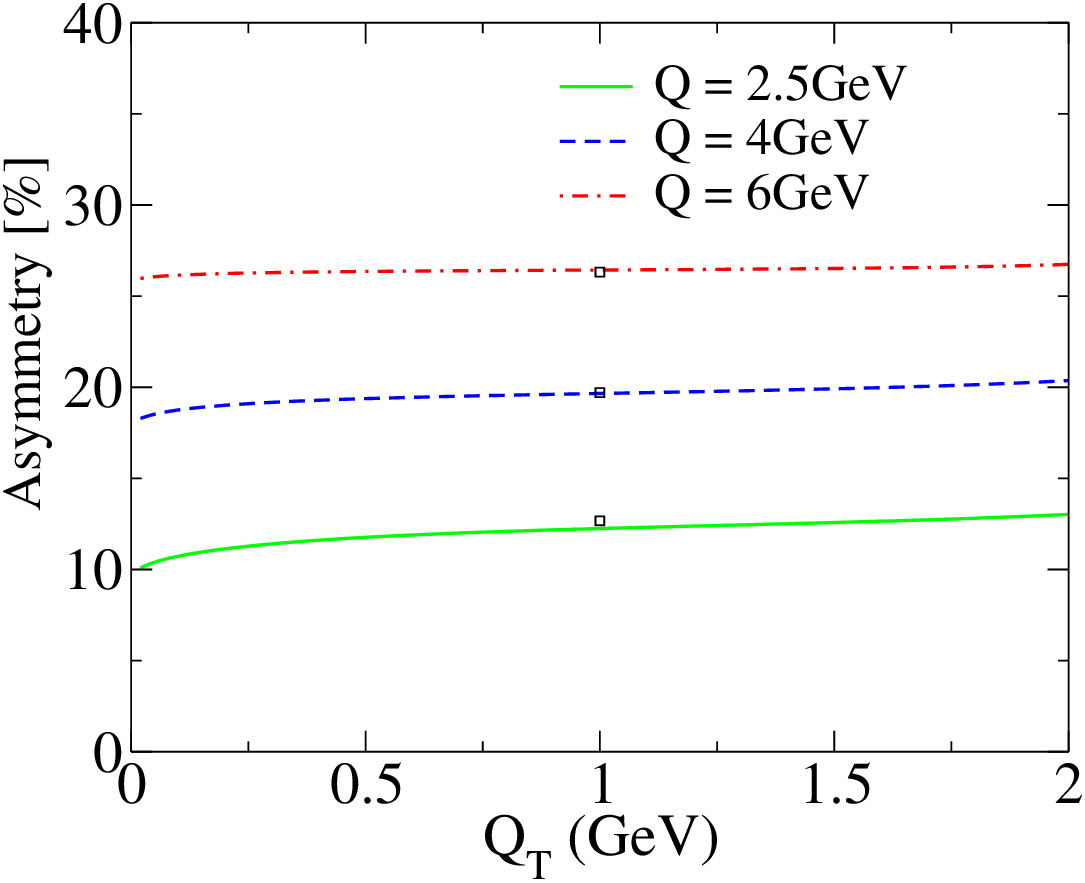}
\ec
\caption{Same as Fig.~4(b), but for
$S=80$ GeV$^2$.
}
\end{figure}

The saddle-point formula (\ref{eq:attnll}) has a particularly simple structure, 
including only the parton distribution functions at the fixed values 
$x=x_{1,2}^0$ and 
at a single scale $b_0/b_{SP}\simeq 1$~GeV, and, combined with (\ref{fullNLL}), 
allows us to obtain $\aqt$ at $Q_T \simeq 1$ GeV.
In Table~1, the row labeled ``SP'' lists the results\footnote{
These results correspond to ``SP-II''
discussed in \cite{KKT:07,KKT:07-2}.}
obtained using the saddle-point formula (\ref{eq:attnll}), with the 
kinematics and nonperturbative inputs of Figs.~4 (a) and (b);
for convenience, the corresponding values are plotted by the symbol ``{\tiny $\Box$}''
at $Q_T=1$ GeV in Figs.~4 (a) and (b). 
``NB'' in Table~1
lists $\aqtn^{\rm NLL}(Q_T=0)$ obtained from (\ref{asymNLL})
using the numerical $b$-integration of (\ref{resum}) and (\ref{UNPOL}).
We observe that the simple formula (\ref{eq:attnll}) has indeed the remarkable accuracy,
reproducing the results of NB as well as the NLL+LO $\aqt$ to 10\% accuracy, i.e.,
to the canonical size of ${\cal O}(\alpha_s)$ corrections 
associated with the NLL accuracy 
(see the discussion below (\ref{eq:speval})).
We also emphasize that the saddle-point formula (\ref{eq:attnll})
indicates that the $Q$ dependence of $\aqt$ is controlled by the detailed
$x_{1,2}^0$ dependence of 
the transversity and density distributions at the scale $b_0/b_{SP}$.
Namely the $Q$ dependence
in Fig.~4 can be understood 
by 
the behavior of the solid lines in Fig.~3 
using (\ref{valence2}).

In Fig.~5, we show the NLL+LO asymmetry $\aqt$ of (\ref{asym}) 
at an another kinematics in the collider mode, $S=80$ GeV$^2$ and $y=0$, 
with the transversity distributions
using (\ref{input}). The results are displayed similarly as Fig.~4.
The observed pattern is similar as Fig.~4(b), but the asymmetries are 
considerably larger.

Figure~6 shows the NLL+LO asymmetries $\aqt$ 
in the fixed-target mode, which is associated with
more challenging kinematic regime for the application of QCD factorization framework:
(a) and (b) plot the results for $S=30$ and 45 GeV$^2$, respectively, 
with $y=0$ using (\ref{input}). 
The dot-dashed curve in (a) ends at the kinematical 
upper bound for the partonic subprocess, 
$Q_{T,\rm{max}}=Q\sqrt{[1-(x_{1}^0)^2][1-(x_{2}^0)^2]}/(x_1^0 +x_2^0) =1.3$ GeV;
because of this kinematical bound, 
we do not show the cases with $Q$ higher than $4$~GeV.
Compared to (a), the case (b) probes smaller $x_{1,2}^0$ region, 
resulting in the smaller asymmetries.
Both results give larger asymmetries than the corresponding collider 
results with the same $Q$.
The characteristic behaviors as functions of $Q_T$ and $Q$ emphasized above
are again observed, and
actually the specific contributions associated with the results in Fig.~6
obey similar pattern as in Figs.~2~(a) and (b), with e.g.,
$\aqt \simeq \aqtn^{\rm NLL}(Q_T) \simeq \aqtn^{\rm LL}$.

\begin{figure}[!t]
\bc
\includegraphics[height=5.8cm]{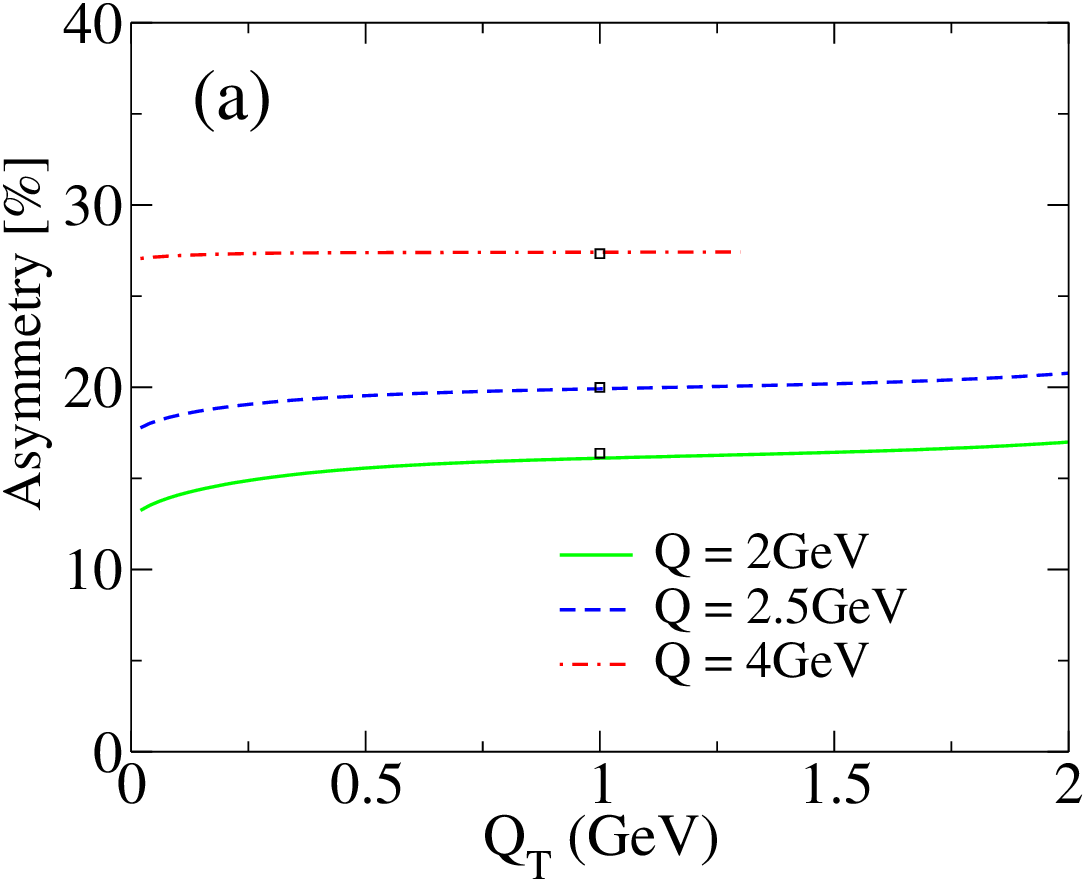}~~~~
\includegraphics[height=5.8cm]{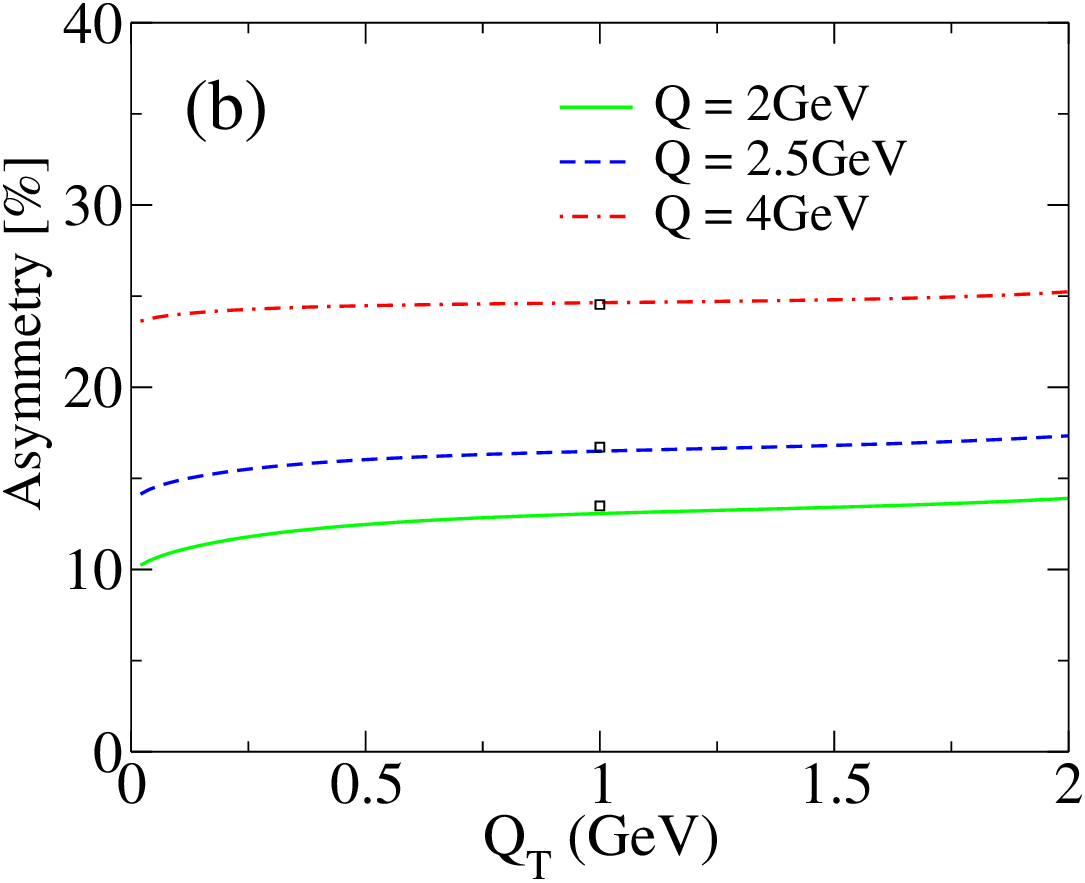}
\ec
\caption{Same as Fig.~4(b), but for 
(a) $S=30$ GeV$^2$ and (b) $S=45$ GeV$^2$. 
}
\end{figure}

We mention the rapidity dependence of the NLL+LO 
asymmetry (\ref{asym}).
Figures~7~(a) and (b) are same as Figs.~4(b) and 6(b) 
in the collider and fixed-target modes, respectively, but for $y=0.5$ and $0.3$,
and all results agree nicely with 
the saddle-point results using (\ref{eq:attnll}).
The characteristic behaviors common to the preceding results are again observed,
but the values of asymmetry are slightly smaller than the corresponding
values for $y=0$.

Figure~8
summarizes our results for the double transverse-spin asymmetry $\aqt$ 
of (\ref{asym}) at GSI as functions of the dilepton mass $Q$ with $y=0$ 
and various values of $S$ corresponding to the fixed-target as well as collider mode.
The symbols ``{\scriptsize $\bigtriangleup$}'' and ``{\scriptsize $\bigtriangledown$}''
plot $\aqtn (Q_T = Q_{T, {\rm peak}})$ using the NLO transversity distributions
corresponding to the Soffer bound and (\ref{input}), respectively,
where $Q_{T, {\rm peak}}$ denotes the value of $Q_T$ at which 
the peak of the corresponding NLL+LO tDY cross section is located (see Fig.~2(a));
e.g., the several symbols ``{\scriptsize $\bigtriangledown$}'' in 
Figs.~8~(a), (b), (c), and (d) denote the values of the 
curves at $Q_T=Q_{T, {\rm peak}}$ ($\simeq 1$ GeV) in Figs.~6(a), 6(b), 5 and 4(b), 
respectively.
The dashed and dot-dashed curves draw the results of the saddle-point 
formula~(\ref{eq:attnll}) using the same parton distributions as those
for ``{\scriptsize $\bigtriangleup$}'' and ``{\scriptsize $\bigtriangledown$}'', 
respectively. 
The results remind the readers of the relation (\ref{fullNLL}). 
Combined with the result in Fig.~3, this relation convinces us that the property,
$\aqt \simeq \aqtn^{\rm NLL}(Q_T) \simeq \aqtn^{\rm LL}$, which was discussed 
below (\ref{valence2}), indeed holds at all relevant kinematics at GSI.

\begin{figure}
\bc
\includegraphics[height=5.8cm]{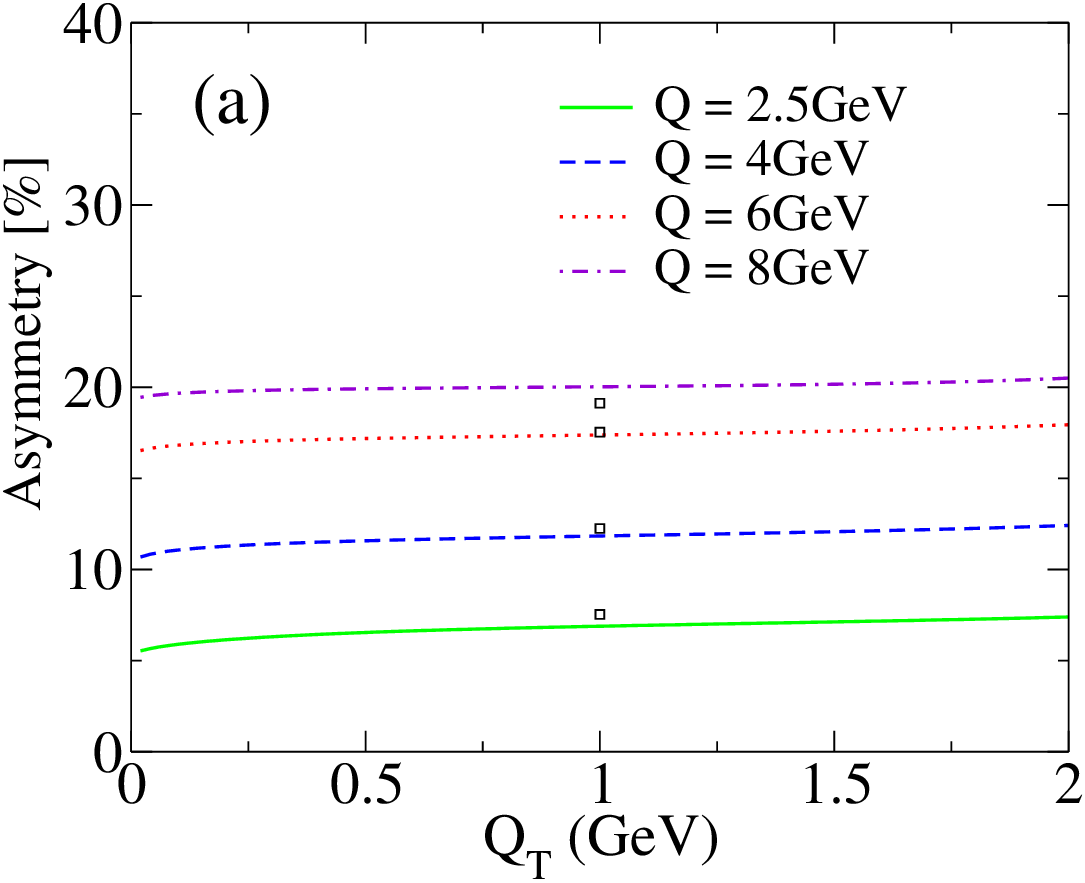}~~~~
\includegraphics[height=5.8cm]{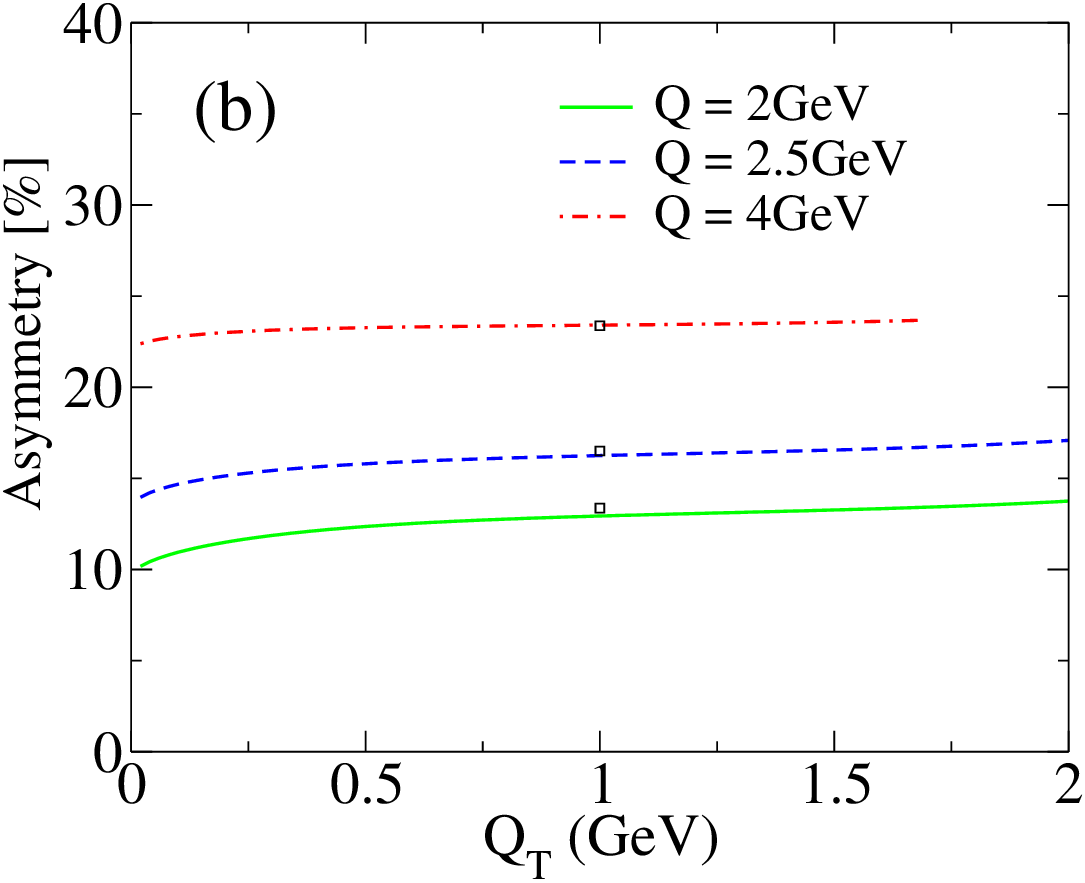}
\ec
\caption{(a) Same as Fig.~4(b), but for $y=0.5$. (b) Same as Fig.~6(b),
 but for $y=0.3$.}
\end{figure}

The conventional double transverse-spin asymmetry $A_{TT}$ is defined 
as the ratio of the $Q_T$-integrated polarized and unpolarized DY cross 
sections, and $A_{TT}$ at GSI kinematics has been studied in
previous works~\cite{Anselmino:2004ki,BCCGR:06,SSVY:05}.
It is worth noting the relation of $A_{TT}$ with our $Q_T$-dependent asymmetry $\aqt$.
As explicitly demonstrated in \cite{KKT:07-2}, the integral of (\ref{NLL+LO}) 
over $Q_T$ reproduces {\it exactly} the corresponding NLO cross sections 
for the $Q_T$-unobserved case;
indeed
$\lambda = 0$ at $b=0$ in (\ref{eq:lambda})
ensures~\cite{KKST:06,KKT:07,BCDeG:03} that the contributions of 
${\cal O}(\alpha_s^2)$ or higher from the resummed 
components (\ref{resum}), (\ref{UNPOL}) vanish in the $Q_T$ integral.
As a result, 
\begin{equation}
A_{TT}\equiv\frac{\int dQ_T^2 \left( \Delta_T d\sigma /dQ^2dQ_T^2dyd\phi \right)}
{\int dQ_T^2 \left( d\sigma /dQ^2dQ_T^2dyd\phi \right)}
=\frac{1}{2}\cos(2 \phi) \frac{\delta H(x_1^0, x_2^0; Q^2)+\cdots}
{H(x_1^0, x_2^0; Q^2)+\cdots}\ ,
\label{eq:att}
\end{equation}
where the ellipses stand for the NLO correction terms, coincides completely
with the NLO asymmetry calculated in \cite{BCCGR:06}.
It has been found~\cite{BCCGR:06} that the effects of the ellipses in (\ref{eq:att}) 
on $A_{TT}$ are generally small at GSI; i.e., 
the corresponding $K$ factors of the polarized and unpolarized cross sections 
are similar to each other and cancel out in the ratio.
Indeed the LO contribution shown explicitly in (\ref{eq:att}) coincides 
with (\ref{asymLL}), i.e., $A_{TT}^{\rm LO} =\aqtn^{\rm LL}$, 
while the NLO calculation yields e.g., $A_{TT}=25.0\%$ for the case of Fig.~2(b). 
\begin{figure}
[!t]
\bc
\includegraphics[width=15cm]{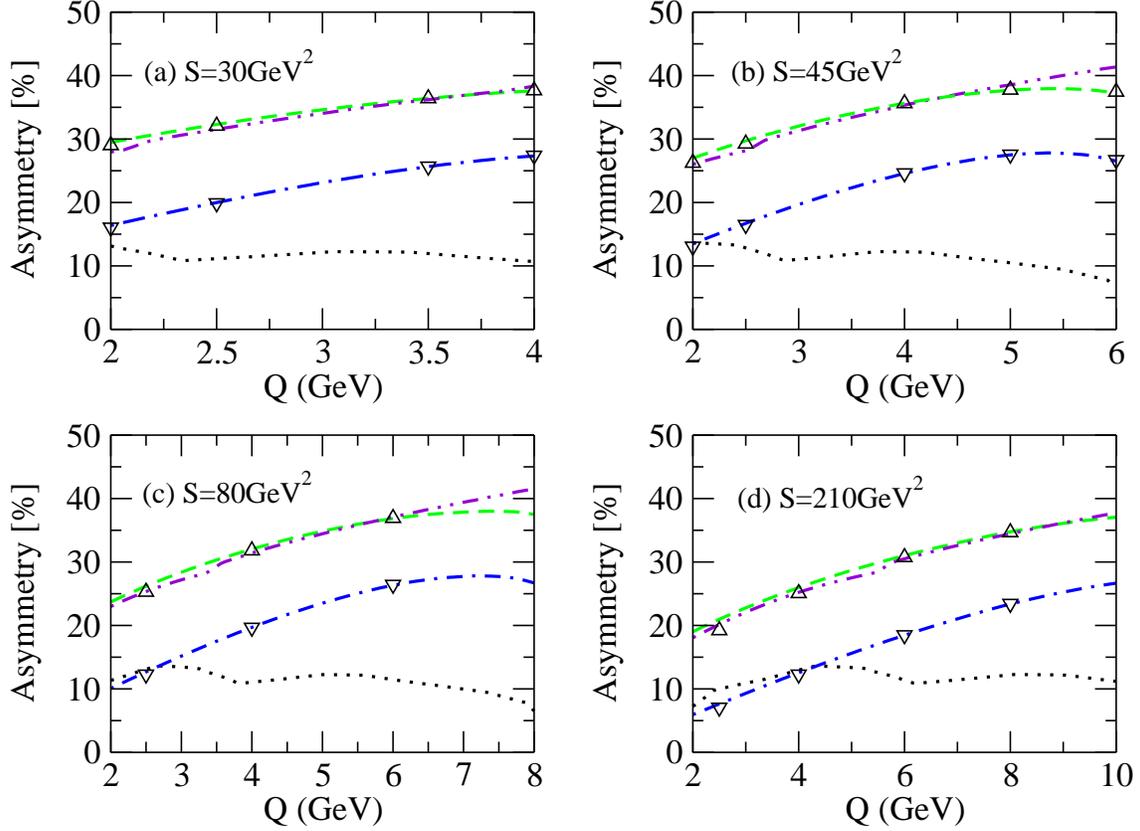}
\ec
\caption{The double transverse-spin asymmetries 
as functions of $Q$ 
at $y=0$.
The dashed and dot-dashed curves plot the results of the saddle-point
formula (\ref{eq:attnll}), using the NLO transversities corresponding to
the Soffer bound and (\ref{input}), respectively, 
which yield the values of
$\aqt$ of (\ref{asym}) for $Q_T = Q_{T,{\rm peak}}$ ($\simeq 1$ GeV) 
as indicated by the triangle up and down symbols.
The two-dot-dashed curve shows the conventional $Q_T$-independent 
LO asymmetry $A_{TT}^{\rm LO}$ using the LO transversities corresponding to 
the Soffer bound. The dotted curve shows 
the ``upper bound'' of $A_{TT}^{\rm LO}$ implied by the result of  
the global fit in \cite{Anselmino:07}. 
}
\end{figure}

In view of the fact $A_{TT} \simeq A_{TT}^{\rm LO}$ at GSI,
it is also interesting to calculate the conventional LO asymmetry $A_{TT}^{\rm LO}$ 
using the LO transversity distributions, obtained by Anselmino 
et al.~\cite{Anselmino:07} through the global fit to the data. 
(For the unpolarized distributions in the denominator of $A_{TT}^{\rm LO}$, 
we use LO GRV98 distributions~\cite{GRV:98}.)
The upper limit of the one-sigma error bounds of their fitted transversities 
(shaded area in Fig.~1)
yields the ``upper bound'' of the corresponding $A_{TT}^{\rm LO}$ as shown
by the dotted curves in Fig.~8. Also plotted by the two-dot-dashed curves 
are $A_{TT}^{\rm LO}$ using the LO transversity distributions corresponding 
to the Soffer bound, which are given by the dot-dashed curves in 
Fig.~1. We here note that the above mentioned relations, 
$\aqt \simeq \aqtn^{\rm NLL}(Q_T) \simeq \aqtn^{\rm LL}$ and 
$A_{TT}^{\rm LO} =\aqtn^{\rm LL}$, imply $\aqt \simeq A_{TT}^{\rm LO}$. 
Therefore, the dotted curves in Fig.~8 represent not only estimate of $A_{TT}$,
but also that of $\aqtn (Q_T \sim Q_{T, {\rm peak}})$, 
using the empirical information of the transversities available at present,
up to anticipated modification of the empirical transversities at the NLO level.
The effect of the corresponding NLO-level modification of the input transversities
on the asymmetries might not be so large, 
as suggested by the fact that the dashed curve in each panel of Fig.~8 is close to
the corresponding two-dot-dashed curve.
In the small $Q$ region, our full result of $\aqt$ using (\ref{input}) can be consistent
with estimate using the empirical LO transversities, but these results have rather 
different behavior
for increasing $Q$, which reflects the different $x$-dependence of 
the corresponding transversities shown in Fig.~1.
This demonstrates that the experimental data to be observed at GSI, 
in particular the behavior 
of the $Q_T$-dependent as well as $Q_T$-independent
asymmetries as functions of $Q$, will allow us to determine
the detailed shape of the transversity distributions in the valence region.

To summarize, the double transverse-spin asymmetry at a measured $Q_T$ to be 
observed 
at GSI is 
very useful to determine
the transversity, and is complementary to the conventional asymmetry associated with 
the $Q_T$-integrated cross sections: 
both asymmetries
are large at GSI, and 
actually the values of these two asymmetries are almost the same for all GSI kinematics.
This (approximate) ``equality'' of the two asymmetries is unexpected 
from the outset and characteristic of $p\bar{p}$ collisions in GSI experiments, and
we have revealed nontrivial roles played by the soft-gluon-resummation contributions. 
Indeed, only after performing
the soft gluon resummation, 
we obtain the ``physical'' behavior for the $Q_T$ spectra of the DY lepton pair
and the associated asymmetry $\aqt$ in the small $Q_T$ region,
with a well-developed peak for the former and the flat behavior for the latter,
and thus we are able to make the reliable estimate.
Furthermore, the resummation modifies 
the parton distributions involved in $\aqt$ 
into those with the ``effective'' scale around $Q_T \sim 1$~GeV,
instead of $Q$ 
in
the conventional asymmetry.
Another reason for
the above ``equality'' 
is 
the similarity of the QCD evolution between the transversity and unpolarized 
quark distributions in the valence region relevant to
the GSI kinematics.
These mechanisms have been explicitly embodied by
the novel saddle-point formula,
which relates the asymmetry $\aqt$ 
with the transversity distributions at the scale around $1$~GeV,
in a way as simple as in the conventional LO asymmetry.
Thus GSI measurements of the asymmetries
at a small $Q_T$ for a variety of dilepton mass $Q$ 
directly probe the shape of the transversity distributions.

\section*{Acknowledgments}
We thank Werner Vogelsang, Hiroshi Yokoya and Stefano Catani
for valuable
discussions. 
We also thank 
Mauro Anselmino and Alexei Prokudin for helpful discussions
on their results in \cite{Anselmino:07}.
This work was supported by the Grant-in-Aid 
for Scientific Research No.
B-19340063.

\end{document}